\documentclass[runningheads]{llncs}

\usepackage{mathpartir}
\usepackage{mathtools}
\usepackage{listings}
\usepackage{color}
\usepackage{framed}
\usepackage{xspace}
\usepackage{caption}
\usepackage{latexsym}

\definecolor{pblue}{rgb}{0.13,0.13,1}
\definecolor{pgreen}{rgb}{0,0.5,0}
\definecolor{pred}{rgb}{0.9,0,0}
\definecolor{pgrey}{rgb}{0.46,0.45,0.48}
\definecolor{javapurple}{rgb}{0.5,0,0.35}

\usepackage{color}
\usepackage{tikz}
\usetikzlibrary{calc,trees,positioning,arrows,chains,shapes.geometric,%
	decorations.pathreplacing,decorations.pathmorphing,shapes,%
	matrix,shapes.symbols,automata}
\usepackage{pgf}

\tikzset{
	>=stealth',
	punktchain/.style={
		rectangle, 
		rounded corners, 
		draw=black, very thick,
		text width=15.5em, 
		minimum height=2em, 
		text centered, 
		on chain},
	line/.style={draw, thick, <-},
	element/.style={
		tape,
		top color=white,
		bottom color=blue!50!black!60!,
		minimum width=8em,
		draw=blue!40!black!90, very thick,
		text width=10em, 
		minimum height=3.5em, 
		text centered, 
		on chain},
	every join/.style={->, thick,shorten >=1pt},
	decoration={brace},
	tuborg/.style={decorate},
	tubnode/.style={midway, right=2pt},
}

\usepackage{balance}
\usepackage{microtype}
\usepackage{enumitem}
\newtheorem{Rule}{Refinement Rule}

\newcommand{\lub}{\ensuremath{\mathit{lub}}}
\newcommand{\mdf}{\ensuremath{\mathit{mdf}}}
\newcommand{\security}{\ensuremath{\mathit{sec}}}
\newcommand{\class}{\ensuremath{\mathit{class}}}
\newcommand{\methTypes}{\ensuremath{\mathit{methTypes}}}
\newcommand{\fields}{\ensuremath{\mathit{fields}}}

\newcommand{\C}{\ensuremath{\mathit{C}}}

\renewcommand{\P}{\ensuremath{\mathit{P}}}

\newcommand{\Q}{\ensuremath{\mathit{Q}}}

\newcommand{\eA}{\ensuremath{\mathit{eA}}}
\newcommand{\m}{\ensuremath{\mathit{m}}}
\newcommand{\s}{\ensuremath{\mathit{s}}}
\newcommand{\T}{\ensuremath{\mathit{T}}}
\newcommand{\mut}{\ensuremath{\mathtt{mut}}}

\newcommand{\imm}{\ensuremath{\mathtt{imm}}}

\renewcommand{\read}{\ensuremath{\mathtt{read}}}
\newcommand{\capsule}{\ensuremath{\mathtt{capsule}}}
\newcommand{\high}{\ensuremath{\mathtt{high}}}
\newcommand{\low}{\ensuremath{\mathtt{low}}}

\newcommand{\Card}{\ensuremath{\mathtt{Card}}}
\newcommand{\Balance}{\ensuremath{\mathtt{Balance}}}
\newcommand{\Pin}{\ensuremath{\mathtt{Pin}}}

\newcommand{\intTT}{\ensuremath{\mathtt{int}}}

\newcommand{\blc}{\ensuremath{\mathtt{blc}}}
\newcommand{\numberTT}{\ensuremath{\mathtt{number}}}
\newcommand{\client}{\ensuremath{\mathtt{client}}}
\newcommand{\emailKw}{\ensuremath{\mathtt{email}}}

\newcommand{\declassify}{\ensuremath{\mathtt{declassify}}}

\newcommand{\mdfarrow}{\ensuremath{\rhd}}

\newcommand{\E}{\ensuremath{\mathcal{E}}}
\newcommand{\classKw}{\text{\footnotesize\ttfamily\bfseries\color{javapurple}class}}
\newcommand{\methodKw}{\text{\footnotesize\ttfamily\bfseries\color{javapurple}method}}
\newcommand{\interfaceKw}{\text{\footnotesize\ttfamily\bfseries\color{javapurple}interface}}
\newcommand{\returnKw}{\text{\footnotesize\ttfamily\bfseries\color{javapurple}return}}
\newcommand{\implementsKw}{\text{\footnotesize\ttfamily\bfseries\color{javapurple}implements}}
\newcommand{\newKw}{\text{\footnotesize\ttfamily\bfseries\color{javapurple}new}}
\newcommand{\thisKw}{\text{\footnotesize\ttfamily\bfseries\color{javapurple}this}}
\newcommand{\extendsKw}{\text{\footnotesize\ttfamily\bfseries\color{javapurple}extends}}
\newcommand{\MOne}{\ensuremath{\mathit{M_1}}}

\newcommand\lst\lstinline

\usepackage{xcolor}
\newcommand\hl[1]{\textcolor{black}{#1}}

\newcommand{\IFbCOO}{IFbCOO\xspace}

\usepackage{graphicx}
\usepackage[colorinlistoftodos]{todonotes}

\definecolor{javapurple}{rgb}{0.5,0,0.35}
\definecolor{key-color}{rgb}{0.8, 0.47, 0.196}

\usepackage{multirow}
\usepackage{multicol}
\usepackage{booktabs}
\usepackage{tabularx}
\newcolumntype{b}{X}
\newcolumntype{n}{>{\hsize=.5\hsize}X}
\newcolumntype{s}{>{\hsize=.2\hsize}X}
\newcolumntype{Y}{>{\centering\arraybackslash}X}
\makeatletter
\newcommand\cellwidth{\TX@col@width}
\makeatother
\makeatletter
\let\tx@\TX@endtabularx
\def\restoretx{\let\TX@endtabularx\tx@}
\makeatother

\usepackage{stmaryrd}
\usepackage{ltablex}

\usepackage[framemethod=TikZ]{mdframed}

\begin{document}
\renewcommand{\thelstlisting}{\arabic{lstlisting}}
	
	\title{Information Flow Control-by-Construction for an Object-Oriented Language
		Using Type Modifiers}
	
	\author{Tobias Runge\inst{1,2} \textsuperscript{[0000-0002-9154-7743]}
		\and Alexander Kittelmann\inst{1,2} \textsuperscript{[0000-0002-8804-7051]}
		\and Marco Servetto\inst{3}
		\and Alex Potanin\inst{4} \textsuperscript{[0000-0002-4242-2725]}
		\and Ina Schaefer\inst{1,2}
	}
	
	\institute{TU Braunschweig, Braunschweig, Germany
		\and Karlsruhe Institute of Technology, Karlsruhe, Germany
		\and Victoria University of Wellington, Wellington, New Zealand
		\and Australian National University, Canberra, Australia\\
		\email{\{tobias.runge,alexander.kittelmann,ina.schaefer\}@kit.edu,
					marco@ecs.vuw.ac.nz, alex.potanin@anu.edu.au}
	}
	
	\authorrunning{Tobias Runge et al.}

\maketitle

\begin{abstract}
	In security-critical software applications, confidential information must be prevented from leaking to unauthorized sinks. Static analysis techniques are widespread to enforce a secure information flow by checking a program after construction.
	A drawback of these systems is that incomplete programs during construction cannot be checked properly. The user is not guided to a secure program by most systems. We introduce \IFbCOO, an approach that guides users incrementally to a secure implementation by using refinement rules.
	In each refinement step, confidentiality or integrity (or both) is guaranteed alongside the functional correctness of the program, such that insecure programs are declined by construction. 
	In this work, we formalize \IFbCOO and prove soundness of the refinement rules. We implement \IFbCOO in the tool CorC and conduct a feasibility study by successfully implementing case studies.
\end{abstract}


\section{Introduction}

For security-critical software, it is important to ensure \emph{confidentiality} and \emph{integrity} of data, otherwise attackers could gain access to this secure data.
For example, in a distributed system, one client \texttt{A} has a lower privilege (i.e., a lower security level) than another client \texttt{B}. When both clients send information to each other, security policies can be violated. If \texttt{A} reads secret data from \texttt{B}, confidentiality is violated. If \texttt{B} reads untrusted data from \texttt{A}, the integrity of \texttt{B}'s data is no longer guaranteed.
To ensure security in software, mostly static analysis techniques are used, which check the software after development~\cite{SabelfeldM03}. A violation of security is only revealed after the program is fully developed. If violations occur, an extensive and repetitive repairing process of writing code and checking the security properties with the analysis technique is needed. 
An alternative is to check the security with
language-based techniques such as type systems~\cite{SabelfeldM03} during the development. In such a secure type system, every expression is assigned to a type, and a set of typing rules checks that the security policy is not violated~\cite{SabelfeldM03}. If violations occur, an extensive process of debugging is required until the code is type-checked.

To counter these shortcomings, we propose a constructive approach to directly develop functionally correct programs that are secure by design without the need of a \emph{post-hoc} analysis. Inspired by the correctness-by-construction (CbC) approach for functional correctness~\cite{kourie2012correctness}, we start with a security specification and refine a high-level abstraction of the program stepwise to a concrete implementation using a set of refinement rules. Guided by the security specification defining the allowed security policies on the used data, the programmer is directly informed if a refinement is not applicable because of a prohibited information flow. With \IFbCOO (Information Flow control by Construction for an Object-Oriented language), programmers get a local warning as soon as a refinement is not secure, which can reduce debugging effort.
With \IFbCOO, functionally correct and secure programs can be developed because both, the CbC refinement rules for functional correctness and the proposed refinement rules for information flow security, can be applied simultaneously.

In this paper, we introduce \IFbCOO which supports information flow control for an object-oriented language with type modifiers for mutability and alias control~\cite{giannini2019flexible}. 
\IFbCOO is based on IFbC~\cite{runge2020lattice} proposed by some of the authors in previous work, 
but lifts its programming paradigm from a simple imperative language to an object-oriented language. 
IFbC introduced a sound set of refinement rules to create imperative programs following an information flow policy, but the language itself is limited to a simple while-language. In contrast, \IFbCOO is based on the secure object-oriented language SIFO~\cite{runge2022sifo}. SIFO's type system uses immutability and uniqueness properties to facilitate information flow reasoning.
In this work, we translate  SIFO's typing rules to refinement rules as required by our correctness-by-construction approach.
This has the consequence that programs written in SIFO and programs constructed using \IFbCOO are interchangeable. In summary, our contributions are the following.
We formalize \IFbCOO and establish 13 refinement rules. We prove soundness that programs constructed with \IFbCOO are secure. Furthermore, we implement \IFbCOO in the tool CorC and conduct a feasibility study.


\section{Object-Oriented Language SIFO by Example}
\label{sec:sifo}

SIFO~\cite{runge2022sifo} is an object-oriented language that ensures secure information flow through a type system with precise uniqueness and (im)mutability reasoning.
\hl{SIFO introduces four type modifiers for references, namely \read, \mut, \imm, and \capsule, which define allowed aliasing and mutability of objects in programs. While, \mut\ and \imm\ point to mutable and immutable object respectively, a \capsule\ reference points to a mutable object that cannot be accessed from other \mut\ references. A \read\ reference points to an object that cannot be aliased or mutated.
In this section, SIFO is introduced with examples to give an overview of the expressiveness and the security mechanism of the language.  We use in the examples two security levels, namely \low\ and \high. An information flow from \low\ to \high\ is allowed, whereas the opposite flow is prohibited. The security levels can be arranged in any user-defined lattice.}
In Section~\ref{sec:formalizing}, we introduce SIFO formally.
In Listing~\ref{code:decl}, we show the implementation of a class \Card\ containing a \low\ immutable \intTT\ \numberTT\ and two \high\ fields: 
a mutable \Balance\ and an immutable \Pin.

\begin{lstlisting}[caption={Class declarations},captionpos=b,label={code:decl}]
class Card{low imm int number; high mut Balance blc;
  high imm Pin pin;}
class Balance{low imm int blc;}
class Pin{low imm int pin;}
\end{lstlisting}

In Listing~\ref{code:imm}, we show allowed and prohibited field assignments with immutable objects as information flow reasoning is the easiest with these references. In a secure assignment, the assigned expression and the reference need the same security level (Lines 6,7). This applies to mutable and immutable objects. 
The security level of expressions is calculated by the least upper bound of the accessed field security level and the receiver security level.
A \high\ \intTT\ cannot be assigned to a \low\ \blc\ reference (Line 8) because this would leak confidential information to an attacker, when the attacker reads the \low\ \blc\ reference. The assignment is rejected.
Updates of a \high\ immutable field are allowed with a \high\ \intTT\ (Line 9) or with a \low\ \intTT\ (Line 10). The \imm\ reference guarantees that the assigned integer is not changed, therefore, no new confidential information can be introduced and a promotion in Line 10 is secure. The promotion alters the security level of the assigned expression to be equal to the security level of the reference.
As expected, the opposite update of a \low\ field with a \high\ \intTT\ is prohibited in Line 11 because of the direct flow from higher to lower security levels.


\begin{lstlisting}[caption={Examples with immutable objects},captionpos=b,firstnumber=5,label={code:imm}]
low mut Card c = new low Card();//an existing Card reference
high mut Balance blc = c.blc;//correct access of high blc
high imm int blc = c.blc.blc;//correct access of high blc.blc
low imm int blc = c.blc.blc;//wrong high assigned to low
c.blc.blc = highInt;//correct field update with high int
c.blc.blc = c.number;//correct update with promoted imm int
high imm int highInt = 0;//should be some secret value
c.number = highInt;//wrong, high int assigned to low c.number
\end{lstlisting}

Next, in Listing~\ref{code:mut}, we  exemplify which updates of mutable objects are legal and which updates are not. We have a strict separation of mutable objects with different security levels. We want to prohibit that an update through a higher reference is read by lower references, or that an update through lower references corrupt data of higher references. A new \Balance\ object can be initialized as a \low\ object because the \Balance\ object itself is not confidential (Line 12). The association to a \Card\ object makes it a confidential attribute of the \Card\ class. However, the assignment of a \low\ \mut\ object to a \high\ reference is prohibited. If Line 13 would be accepted, Line 14 could be used to insecurely update the confidential \Balance\ object because the \low\ reference is still in scope of the program. Only an assignment without aliasing is allowed (Line 16). With \capsule, an encapsulated object is referenced to which no other \mut\ reference points. The \low\ \lst@capsBlc@ object can be promoted to a \high\ security level and assigned. Afterwards, the \capsule\ reference is no longer accessible.
In the case of an immutable object, the aliasing is allowed (Line 18), since the object itself cannot be updated (Line 19). Both \imm\ and \capsule\ references are usable to communicate between different security levels.

\begin{lstlisting}[caption={Examples with mutable and encapsulated objects},captionpos=b,firstnumber=12,label={code:mut}]
low mut Balance newBlc = new low Balance(0);//ok
c.blc = newBlc;//wrong, mutable secret shared as low and high
newBlc.blc = 10;//ok? Insecure with previous line
low capsule Balance capsBlc = new low Balance(0);//ok
c.blc = capsBlc;//ok, no alias introduced
low imm Pin immPin = new low Pin(1234);//ok
c.pin = immPin;//ok, pin is imm and can be aliased
immPin.pin = 5678;//wrong, immutable object cannot be updated
\end{lstlisting}

\section{\IFbCOO by Example}

With \IFbCOO, programmers can incrementally develop programs, where the security levels are organized in a lattice structure to guarantee a variety of confidentiality and integrity policies. \IFbCOO defines 13 refinement rules to create secure programs. As these rules are based on refinement rules for correctness-by-construction, programmers can simultaneously apply refinements rules for functional correctness~\cite{kourie2012correctness,runge2019tool,bordis2022recorc} and security. We now explain \IFbCOO in the following examples. For simplicity, we omit the functional specification. \IFbCOO is introduced formally in Section~\ref{sec:formalizing}.

In \IFbCOO, the programmer starts with a class including fields of the class and declarations of method headers. \IFbCOO is used to implement methods in this class successively. The programmer chooses one abstract method body and refines this body to a concrete implementation of the method.
A starting \IFbCOO tuple specifies the typing context $\Gamma$ and the abstract method body $\eA$. The expression $\eA$ is abstract in the beginning and refined incrementally to a concrete implementation.
During the construction process, local variables can be added. The refinement process in \IFbCOO results in a method implementation which can be exported to the existing class.
First, we give a fine-grained example to show the application of refinement rules in detail. The second example illustrates that \IFbCOO can be used to implement larger methods.

The first example in Listing~\ref{code:set} is a setter method. A field $\mathtt{number}$ is set with a parameter $\mathtt{x}$.
We start the construction with an abstract expression $\eA : [\Gamma;\low\ \imm\ \mathtt{void}]$ with a typing context $\Gamma = \low\ \mut\ C\ \mathtt{this}, \low\ \imm\ \mathtt{int}\ \mathtt{x}$ extracted from the method signature ($C$ is the class of the method receiver). 
The abstract expression $\eA$ contains all local information (the typing context and its type) to be further refined. A concrete expression that replaces the abstract expression must have the same type $\low\ \imm\ \mathtt{void}$, and it can only use variables from the typing context $\Gamma$.
The tuple $[\Gamma;\low\ \imm\ \mathtt{void}]$ is now refined stepwise. First, we introduce a field assignment: $\eA \rightarrow \eA_1.\mathtt{number} = \eA_2$. The newly introduced abstract expressions are $\eA_1 : [\Gamma;\low\ \mut\ C]$ and $\eA_2 : [\Gamma;\low\ \imm\ \mathtt{int}]$ according to the field assignment refinement rule. In the next step, $\eA_1$ is refined to $\mathtt{this}$, which is the following refinement: $\eA_1.\mathtt{number} = \eA_2 \rightarrow \mathtt{this.number} = \eA_2$. As $\mathtt{this}$ has the same type as $\eA_1$, the refinement is correct. The last refinement replaces $\eA_2$ with $\mathtt{x}$, resulting in $\mathtt{this.number} = \eA_2 \rightarrow \mathtt{this.number} = \mathtt{x}$. As $\mathtt{x}$ has the same type as $\eA_2$, the refinement is correct. The method is fully refined since no abstract expression is left.

\begin{lstlisting}[caption={Set method},captionpos=b,label={code:set}]
low mut method low imm void setNumber(low imm int x) {
  this.number = x; }
\end{lstlisting}

To present a larger example, we construct a check of a signature in an email system (see Listing~\ref{code:example-signature}). The input of the method is an \emailKw\ object and a \client\ object that is the receiver of the email. The method checks whether the key with which the \emailKw\ object was signed and the stored public key of the \client\ object are a valid pair. If this is the case, the \emailKw\ object is marked as verified. The fields $\mathtt{isSignatureVerified}$ and $\mathtt{emailSignKey}$ of the class \emailKw\ have a \high\ security level, as they contain confidential data. The remaining fields have \low\ as security level.

	{\centering
	\begin{minipage}{\linewidth}
\begin{lstlisting}[
		caption={Program of a secure signature verification},
		label={code:example-signature}
		]
static low imm void verifySignature(
  low mut Client client, low mut Email email) { 
  low imm int pubkey = client.publicKey;
  high imm int privkey = email.emailSignKey;
  high imm boolean isVerified;
  if (isKeyPairValid(privkey, pubkey)) {
    isVerified = true;
  } else {
    isVerified = false;
  }
  email.IsSignatureVerified = isVerified;
}
\end{lstlisting}
	\end{minipage}}

\begin{figure}[t]
	\centering
	\includegraphics[width=\linewidth]{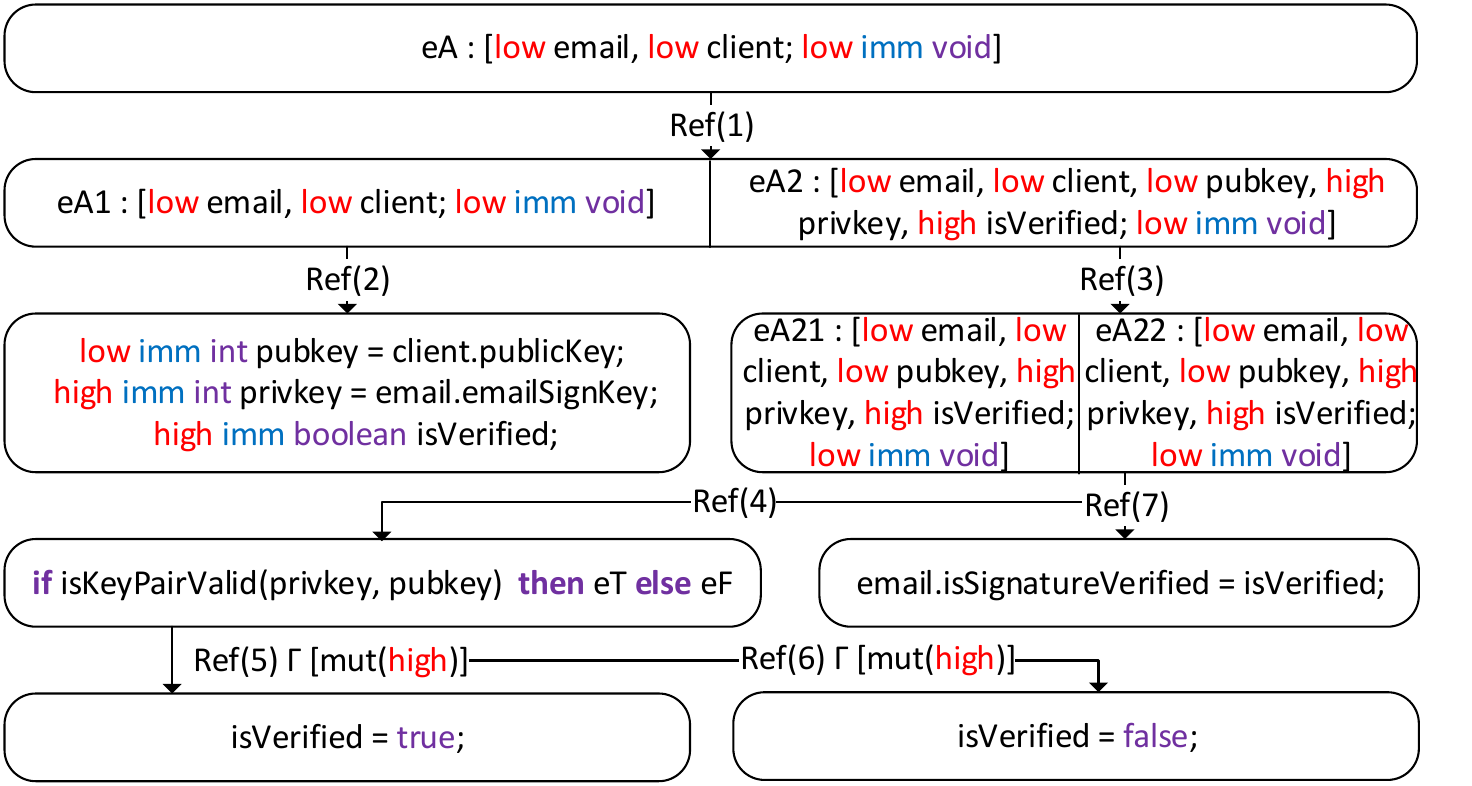}
	\caption{Refinement steps for the signature example}
	\label{fig:refinement}
\end{figure}

In Figure~\ref{fig:refinement}, we show the starting \IFbCOO tuple with the security level of the variables (type modifier and class name are omitted) at the top.
In our example, we have two parameters \client\ and \emailKw, with a \low\ security level. 
To construct the algorithm of Listing~\ref{code:example-signature}, the method implementation is split into three parts. First, two local variables (private and public key for the signature verification) are initialized and a Boolean for the result of the verification is declared. Second, verification whether the keys used for the signature form a valid pair takes place. Finally, the result is saved in a field of the \emailKw\ object.

Using the refinement rule for composition, the program is initially split into the initialization phase and the remainder of the program's behavior (Ref(1)). This refinement introduces two abstract expressions $\eA1$ and $\eA2$. The typing contexts of the expressions are calculated by \IFbCOO automatically during refinement. As we want to initialize two local variables by further refining $\eA1$, the finished refinement in Figure~\ref{fig:refinement} already contains the local \high\ variables $\mathtt{privkey}$ and $\mathtt{isVerified}$, and the \low\ variable $\mathtt{pubkey}$ in the typing context of expression $\eA2$.

In Ref(2), we apply the assignment refinement\footnote{To be precise, it would be a combination of composition and assignment refinements, because an assignment refinement can only introduce one assignment expression.} to initialize the integers $\mathtt{pubkey}$ and $\mathtt{privkey}$. Both references point to immutable objects that are accessed via fields of the objects \client\ and \emailKw. The security levels of the field accesses are determined with the field access rule checked by \IFbCOO. The determined security level of the assigned expression must match the security level of the reference. In this case, the security levels are the same.
Additionally, it is enforced that immutable objects cannot be altered after construction (i.e., it is not possible to corrupt the private and public key). In Ref(3), the next expression $\eA2$ is split with a composition refinement into $\eA21$ and $\eA22$.

Ref(4) introduces an if-then-else-expression by refining $\eA21$. Here, it is checked whether the public and private key pair is valid. As the $\mathtt{privkey}$ object has a \high\ security level, we have to restrict our typing context with $\Gamma[\mathit{mut}(\high)]$. 
This is necessary to prevent indirect information leaks. 
With the restrictions, we can only assign expressions to at least \high\ references and mutate \high\ objects ($\mathit{mut}(\high)$) in the then- and else-expression.
If we assign a value in the then-expression to a \low\ reference that is visible outside of the then-expression, an attacker could deduce that the guard was evaluated to true by reading that \low\ reference.

Ref(5) introduces an assignment of an immutable object to a \high\ reference, which is allowed in the restricted typing context. As explained, the assignment to \low\ references is forbidden. The assigned immutable object $\mathtt{true}$ can be securely promoted to a \high\ security level.
In Ref(6), a similar assignment is done, but with the value $\mathtt{false}$.
Ref(7) sets a field of the \emailKw\ object by refining $\eA22$. We update the \high\ field of the \emailKw\ object by accepting the \high\ expression $\mathtt{isVerified}$.
With this last refinement step, the method is fully concretized. The method is secure by construction and constitutes valid SIFO code (see Listing~\ref{code:example-signature}).

\section{Formalizing Information Flow Control-by-Construction}
\label{sec:formalizing}

In this section, we formalize \IFbCOO for the construction of functionally correct and secure programs. Before, we introduce SIFO as the underlying programming language formally.

\subsection{Core Calculus of SIFO}

\begin{figure}[t]
	\begin{minipage}{0.97\linewidth}
		\begin{array}[t]{lcl}
			\T &::= &\s\ \mdf\ \C\\
			\s &::= &\high\ \vert\ \low\ \vert\ \dots  (\text{user defined})\\
			\mdf &::= &\mut\ \vert\ \imm\ \vert\ \capsule\ \vert\ \read\\
			\mathit{CD} &::= &\text{\lst@class@}\ \C\ \text{\lst@implements@}\ \overline{\C}\ \{\overline{F}\ \overline{MD}\ \}
			{}_{}\ \vert\ \text{\lst@interface@}\ \C\ \text{\lst@extends@}\ \overline{\C}\ \{\overline{\mathit{MH}}\}\\
			F &::= &\s\ \mut\ \C\ f;\ \vert\ \s\ \imm\ \C\ f;\\
			MD &::= &\mathit{MH}\ \{\text{\lst@return@}\ e\text{\lst@;@}\}\\
			\mathit{MH} &::= &\s\ \mdf\ \text{\lst@method@}\ \T\ \m (\T_{1}\ x_{1},\ \dots, \T_{n}\ x_{n})\\
			e &::=& \eA\ \vert\ x\ \vert\ e_0.f=e_1\  \vert\ e.f\ \vert\ e_0.m(\overline{e})\ \vert\ \text{\lst@new@}\ \s\ \C(\overline{e})\
			 \vert\ e_0;e_1\\
			&&
			\vert\ \text{\lst@if@}\ e_0\ \text{\lst@then@} \ e_1\ \text{\lst@else@}\ e_2\
			\vert\ \text{\lst@while@}\ e_0\ \text{\lst@do@} \ e_1\ \vert\ \text{\lst@declassify@}(e)\\
			\Gamma &::= & x_1 : T_1 \dots x_n : T_n\\
			\E &::= &[]\ \vert\ \E.f\ \vert\ \E.f=e\ \vert\ e.f=\E \vert\ \E.m(\overline{e})\ \vert\ e.m(\overline{e}\ \E\ \overline{e})\ \vert\ \text{\lst@new@}\ \s\ \C(\overline{e}\ \E\ \overline{e})\\
		\end{array}
	\end{minipage}
	\caption{Syntax of the extended core calculus of SIFO}
	\label{f:syntax}
\end{figure}

\hl{Figure~\ref{f:syntax} shows the syntax of the extended core calculus of SIFO~\cite{runge2022sifo}. SIFO is an expression-based language similar to Featherweight Java~\cite{igarashi2001featherweight}.}
Every reference and expression is associated with a type \T. The type \T\ is composed of a security level \s, a type modifier \mdf\ and a class name \C. Security levels are arranged in a lattice with one greatest level $\top$ and one least level $\bot$ forming the security policy.
The security policy determines the allowed information flow.
Confidentiality and integrity can be enforced by using two security lattices and two security annotations for each expression. 
Each property is enforced by a strict separation of security levels. In the interest of an expressive language, we allow the information flow from lower to higher levels (confidentiality or integrity security levels) using promotion rules while the opposite needs direct interaction with the programmer by using the \declassify\ expression. For convenience, we will use only one lattice of confidentiality security levels in the explanations.

The type modifier \mdf\ can be \mut, \imm, \capsule, and \read\ with the following subtyping relation. For all type modifier $\mdf: \capsule$ $\leq \mdf, \mdf\leq\read$.
In SIFO, objects are \textit{mutable} or (deeply) \textit{immutable}. 
The reachable object graph (ROG) from a mutable object is composed of mutable and immutable objects, while the ROG of an immutable object can only contain immutable objects.
A \mut\ reference must point to a mutable object; such an object can be aliased and mutated.
An \imm\ reference must point to an immutable object; such an object can be aliased, but not mutated.
A \capsule\ reference points to a mutable object. The object and the mutable objects in its ROG cannot be accessed from other references. As \capsule\ is a subtype of \imm\ and \mut\, the object can be assigned to both.
Finally, a \read\ reference is the supertype that points to an object that cannot be aliased or mutated, but it has no immutability guarantee that the object is not modified by other references.
These modifiers allow us to make precise decisions about the information flow by utilizing immutability or uniqueness properties of objects. For example, an immutable object cannot be altered, therefore it can be securely promoted to a higher security level. For a mutable object, a security promotion is insecure because an update through other references with lower security levels can corrupt the confidential information.

Additionally, the syntax of SIFO contains 
class definitions $CD$ which can be classes or interfaces. An interface has a list of method headers. 
A class has additional fields. 
A field $F$ has a type \T\ and a name, but the type modifier can only be \mut\ or \imm.
A method definition $MD$ consists of a method header and a body.
The header has a receiver, a return type, and a list of parameters.
The parameters have a name and a type \T. The receiver has a type modifier and a security level.
An expression $e$ can be a variable, field access, field assignment, method call, or object construction in SIFO. In the extended version presented in the paper, we also added abstract expressions, sequence of expressions, conditional expression, loop expression, and declassification. With the \declassify\ operator a reverse information flow is allowed. The expression $\eA$ is abstract and typed by $[\Gamma;T]$. Beside the type \T\, a local typing context $\Gamma$ is used to have all needed information to further refine $\eA$.
We require a Boolean type for the guards in the conditional and loop expression. A typing context $\Gamma$ assigns a type $T_i$ to variable $x_i$.
With the evaluation context \E, we define the order of evaluation for the reduction of the system.
The typing rules of SIFO are shown in Appendix~\ref{sec:appendix_typing}.

\subsection{Refinement Rules for Program Construction}

To formalize the \IFbCOO refinement rules, in Figure~\ref{fig:sbc_notions}, we introduce basic notations, which are used in the refinement rules. 

$L$ is the lattice of security levels to define the information flow policy and \lub\ is used to calculate the least upper bound of a set of security levels.
The functional and security specification of a program is defined by an \IFbCOO tuple $\{P;Q;\Gamma;T;\eA\}$.
The \IFbCOO tuple consists of a typing context $\Gamma$, a type $T$, an abstract expression $\eA$, and a functional pre-/postcondition, which is declared in the first-order predicates \P\ and \Q. The abstract expression is typed by $[P;Q;\Gamma;T]$. In the following, we focus on security, so the functional specification is omitted. 

The refinement process of \IFbCOO starts with a method declaration, where the typing context $\Gamma$ is extracted from the arguments and $T$ is the method return type. Then, the user guides the construction process by refining the first abstract expression $\eA$. 
With the notation
$\Gamma[\mathit{mut}(s)]$, we introduce a restriction to the typing context. The function $\mathit{mut}(s)$ prevents mutation of mutable objects that have a security level lower than $s$.
When the user chooses the lowest security level of the lattice, the function does not restrict $\Gamma$. The function $\security(\T)$ extracts the security level of a type \T.

\begin{figure}[tb]
	\centering
	\begin{minipage}{0.97\linewidth}
		\begin{array}[t]{rcl}
			L && \text{Bounded upper semi-lattice } (L,\leq) \text{ of security levels}\\
			\lub: \mathcal{P}(L) \to L && \text{Least upper bound of the security levels in } L\\
			\{P;Q;\Gamma;T;\eA\} && \text{Starting \IFbCOO tuple} \\
			\eA:[P;Q;\Gamma;T] && \text{Typed abstract expression \eA}\\
			\Gamma[\mathit{mut}(s)] && \text{Restricted typing context}\\
			\security(\T) = \s && \text{Returns the security level \s\ in type \T}
		\end{array}
	\end{minipage}
	\caption{Basic notations for \IFbCOO} 
	\label{fig:sbc_notions}
\end{figure}


\paragraph{Refinement Rules.}

The refinement rules are used to replace an \IFbCOO tuple $\{\Gamma;T;\eA\}$ with a concrete implementation by concretizing the abstract expression $\eA$. This refinement is only correct if specific side conditions hold. On the right side of the rules, all newly introduced symbols are implicitly existentially quantified.
The rules can introduce new abstract expressions $\eA_i$ which can be refined by further applying the refinement rules.

\begin{Rule}[Variable]\quad
	\\$\eA$ is refinable to $x$ if $\eA : [\Gamma;T]$ and  $\Gamma(x) = T$.
\end{Rule}

\noindent The first \IFbCOO rule introduces a variable $x$, which does not alter the program. It refines an abstract expression to an $x$ if $x$ has the correct type \T.

\begin{Rule}[Field Assignment]\quad
	\\$\eA$ is refinable to $\eA_0.f:=\eA_1$ if $\eA : [\Gamma;T]$ and $\eA_0 : [\Gamma;\s_0\ \mut\ \C_0]$ and $\eA_1 : [\Gamma;s_1\ \mdf\ C]$ and  $\s\ \mdf\ \C\ f \in fields(C_0)$  and $s_1 = \lub(s_0,s)$.
\end{Rule}

\noindent
We can refine an abstract expression to a field assignment if the following conditions hold. The expression $\eA_0$ has to be \mut\ to allow a manipulation of the object. The security level of the assigned expression $\eA_1$  has to be equal to the least upper bound of the security levels of expression $\eA_0$ and the field $f$. The field $f$ must be a field of the class $C_0$ with the type $\s\ \mdf\ \C$. With the security promotion rule, the security level of the assigned expression can be altered.


\begin{Rule}[Field Access]\quad
	\\$\eA$ is refinable to $\eA_0.f$ if $\eA : [\Gamma;\s\ \mdf\ \C]$ and $\eA_0 : [\Gamma;\s_0\ \mdf_0\ \C_0]$ and $\s_1\ \mdf_1\ \C\ f \in fields(C_0)$
	and $\s = \lub(s_0,s_1)$ and $\mdf_0 \mdfarrow \mdf_1 = \mdf$.
	
\end{Rule}

\noindent
We can refine an abstract expression to a field access if a field $f$ exists in the class of receiver $\eA_0$ with the type $\s_1\ \mdf_1\ \C$. 
The accessed value must have the expected type $\s\ \mdf\ \C$ of the abstract expression.
This means, the class name of the field $f$ and $C$ must be the same. Additionally, the security level of the abstract expression $\eA$ is equal to the least upper bound of the security levels of expression $\eA_0$ and field $f$. The type modifiers must also comply.
The arrow between type modifiers is defined as follows. As we allow only \mut\ and \imm\ fields, not all possible cases are defined: 
$\mdf \mdfarrow \mdf' = \mdf''$\\
${}_{}\bullet$ $ \mut \mdfarrow \mdf = \capsule \mdfarrow \mdf = \mdf$\\
${}_{}\bullet$ $ \imm \mdfarrow \mdf = \mdf \mdfarrow \imm = \imm$\\
${}_{}\bullet$ $ \read \mdfarrow \mut = \read$.


\begin{Rule}[Method Call]\quad
	\\$\eA$ is refinable to $\eA_0.m(\eA_{1},\dots, \eA_{n})$ if $\eA : [\Gamma;T]$ and $\eA_0 : [\Gamma;T_0] \dots \eA_n : [\Gamma;T_n]$ and $T_0 \dots T_n \rightarrow T \in \methTypes(\class(T_0),\ m)$
	and $\security(T) \geq \security(T_0)$ and $\mathit{forall}\ i \in \{1,\dots,n\}$ if $mdf(T_i) \in \{\mut,\capsule\}$ then  $\security(T_i) \geq \security(T_0)$.
	
\end{Rule}

\noindent
With the method call rule, an abstract expression is refined to a call to method $m$. 
The method has a receiver $\eA_0$, a list of parameters $\eA_1 \dots \eA_n$, and a return value.
A method with matching definition must exist in the class of receiver $\eA_0$. This method definition is returned by the \methTypes\ function. The function \class\ returns the class of a type \T. The security level of the return type has to be greater than or equal to the security level of the receiver.
This condition is needed because through dynamic dispatch information of the receiver may be leaked if its security level is higher than the security level of the return type. The same applies for \mut\ and \capsule\ parameters. The security level of these parameters must also be greater than or equal to the security level of the receiver.
As the method call replaces an abstract expression $\eA$, the return value must have the same type (security level, type modifier, and class name) as the refined expression.
In Appendix~\ref{sec:methods}, we introduce multiple methods types~\cite{runge2022sifo} to reduce writing effort and increase the flexibility of \IFbCOO. A method can be declared with specific types for receiver, parameters and return value, and other signatures of this method are deduced by applying the transformations from the multiple method types definition, where security level and type modifiers are altered. All these deduced method declarations can be used in the method call refinement rule.



\begin{Rule}[Constructor]\quad
	\\$\eA$ is refinable to  $\mathtt{new}\ s\ C(\eA_1\dots \eA_n)$ if $\eA : [\Gamma;\s\ \mdf\ \C]$ and $\fields(C)=T_1 \ f_1\dots T_n\ f_n$ and $\eA_1 : [\Gamma;T_1[s]] \dots \eA_n : [\Gamma;T_n[\s]]$.
	
\end{Rule}

\noindent
The constructor rule is a special method call. We can refine an abstract expression to a constructor call, where a mutable object of class $C$ is constructed with a security level $s$.
The parameter list $\eA_1 \dots \eA_n$ must match the list of declared fields $f_1 \dots f_n$ in class $C$.
Each parameter $\eA_i$ is assigned to field $f_i$. 
This assignment is allowed if the type of parameter $\eA_i$ is (a subtype of) $T_i[s]$.
$T[s]$ is a helper function which returns a new type whose security level is the least upper bound of $\security(T)$ and $s$. It is defined as:
$\T[\s]$ $= \lub(\s,\s')\ \mdf\ \C$, where $T = \s'\ \mdf\ \C$, defined only if $s' \leq s$ or $s \leq s'$.
By calling a constructor, the security level \s\ can be freely chosen to use parameters with security levels that are higher than originally declared for the fields.
In other words, a security level \s\ is used to initialize lower security fields with parameters of higher security level \s. This results in a newly created object with the security level $s$~\cite{runge2022sifo}.
As the newly created object replaces an abstract expression $\eA$, the object must have the same type as the abstract expression.
If the modifier promotion rule is used (i.e., no mutable input value exist), the object can be assigned to a \capsule\ or \imm\ reference.



\begin{Rule}[Composition]\quad
	\\$\eA$ is refinable to $\eA_0;\eA_1$ if $\eA : [\Gamma;T]$ and $\eA_0 : [\Gamma;T_0]$ and $\eA_1 : [\Gamma;T]$.
\end{Rule}

\noindent With the composition rule, an  abstract expression $\eA$  is refined to two subsequent abstract expression $\eA_0$ and $\eA_1$. The second abstract expression must have the same type \T\ as the refined expression. 


\begin{Rule}[Selection]\quad
	\\$\eA$ is refinable to $\mathtt{if}\ \eA_0\ \mathtt{then}\ \eA_1\ \mathtt{else}\ \eA_2$ if $\eA : [\Gamma;T]$ and $\eA_0 : [\Gamma;\s\ \imm\ $ $\mathtt{Boolean}]$ and
	$\eA_1 : [\Gamma[\mathit{mut}(s)];T]$ and $\eA_2 : [\Gamma[\mathit{mut}(s)];T]$.
	
\end{Rule}

\noindent The selection rule refines an abstract expression to a conditional $\mathtt{if}$-$\mathtt{then}$-$\mathtt{else}$- expression. Secure information can be leaked indirectly as the selected branch may reveal the value of the guard. In the branches, the typing context is restricted. The restricted typing context prevents updating mutable objects with a security level lower than $s$. The security level $s$ is determined by the Boolean guard $\eA_0$. When we add updatable local variables to our language, the selection rule must also prevent the update of local variables that have a security level lower than $s$.


\begin{Rule}[Repetition]\quad
	\\$\eA$ is refinable to $\mathtt{while}\ \eA_0\ \mathtt{do}\ \eA_1$ if $\eA : [\Gamma;T]$ and $\eA_0 : [\Gamma;\s\ \imm\ \mathtt{Boolean}]$ and
	$\eA_1 : [\Gamma[\mathit{mut}(s)];T]$.
	
\end{Rule}

\noindent The repetition rule refines an abstract expression to a $\mathtt{while}$-loop. The repetition rule is similar to the selection rule. For the loop body, the typing context is restricted to prevent indirect leaks of the guard in the loop body. The security level $s$ is determined by the Boolean guard $\eA_0$.


%
%
%


\begin{Rule}[Context Rule]\quad
	\\$\E[\eA]$ is refinable to $\E[e]$ if $\eA$ is refinable to $e$.
\end{Rule}

\noindent The context rule replaces in a context \E\ an abstract expression with a concrete expression, if the abstract expression is refinable to the concrete expression.

\begin{Rule}[Subsumption Rule]\quad
	\\$\eA : [\Gamma;T]$  is refinable to $\eA_1 : [\Gamma;T']$ if $T' \leq T$.
\end{Rule}

\noindent The subsumption rule can alter the type of expressions. An abstract expression that requires a type $T$ can be weakened to require a type $T'$ if the type $T'$ is a subtype of $T$.

\begin{Rule}[Security Promotion]\quad
	\\$\eA : [\Gamma;s\ mdf\  C]$ is refinable to $\eA_1 : [\Gamma;s'\ mdf\ C]$ if $\mdf \in \{\capsule,\imm\}$ and $s' \leq s$.
\end{Rule}

\noindent The security promotion rule can alter the security level of expressions. An abstract expression that requires a security level $s$ can be weakened to require a security level $s'$ if the expression is \capsule\ or \imm. Other expressions (\mut\ or \read) cannot be altered because potentially existing aliases are a security hazard.

\begin{Rule}[Modifier Promotion]\quad
	\\$\eA : [\Gamma;s\ \capsule\ C]$  is refinable to $\eA_1 : [\Gamma[\mut\backslash \read];s\ \mut\ C]$.
\end{Rule}

\noindent
The modifier promotion rule can alter the type modifier of an expression $\eA$. An abstract expression that requires a \capsule\ type modifier can be weakened to require a \mut\ type modifier if all \mut\ references are only seen as \read\ in the typing context. That means, that the mutable objects in the ROG of the expression cannot be accessed by other references. Thus, manipulation of the object is only possible through the reference on $\eA$.

\begin{Rule}[Declassification]\quad
	\\$\eA : [\Gamma;\bot\ \mdf\ C]$  is refinable to $\declassify(\eA_1) : [\Gamma;\s\ mdf\ C]$ if $\mdf \in$ $\{\capsule,\imm\}$.
	
\end{Rule}

\noindent
In our information flow policy, we can never assign an expression with a higher security level to a variable with a lower security level. To allow this assignment in appropriate cases, the \declassify\ rule is used. An expression \eA\ is altered to a $\declassify$-expression with an abstract expression $\eA_1$ that has a security level $s$ if the type modifier is \capsule\ or \imm. A \mut\ or \read\ expression cannot be declassified as existing aliases are a security hazard. \hl{Since we have the security promotion rule, the declassified \capsule\ or \imm\ expression can directly be promoted to any higher security level. Therefore, it is sufficient to use the bottom security level in this rule without restricting the expressiveness.}
For example, the rule can be used to assign a hashed password to a public variable. The programmer has the responsibility to ensure that the use of \declassify\ is secure.

\subsection{Proof of Soundness}

In this subsection and Appendix~\ref{sec:appendix_proof}, we prove that programs constructed with the \IFbCOO refinement rules are secure according to the defined information flow policy. We prove this by showing that programs constructed with \IFbCOO are well typed in SIFO (Theorem~\ref{theorem:NI}). SIFO itself is proven to be secure~\cite{runge2022sifo}. In Appendix~\ref{sec:appendix_proof}, we prove this property for the core language of SIFO, which does not contain composition, selection, and repetition expressions. The SIFO core language is minimal, but using well-known encodings, it can support composition, selection, and repetition (encodings of the Smalltalk~\cite{goldberg1983smalltalk} style support control structures). We also
exclude the declassify operation because this rule is an explicit mechanism to break security in a controlled way.

\begin{theorem}[Soundness of \IFbCOO]\label{theorem:NI}\quad\\
	An expression $e$ constructed with \IFbCOO is well typed in SIFO.
\end{theorem}

%

\section{CorC Tool Support and Evaluation}

\IFbCOO is implemented in the tool CorC~\cite{runge2019tool,bordis2022recorc}. CorC itself is a hybrid textual and graphical editor to develop programs with correctness-by-construction. IFbC~\cite{runge2020lattice} is already implemented as extension of CorC, but to support object-orientation with \IFbCOO a redesign was necessary. Source code and case studies are available at: \url{https://github.com/TUBS-ISF/CorC/tree/CCorCOO}.

%
%

\subsection{CorC for IFbCOO}


For space reasons, we cannot introduce CorC comprehensively. We just summarize the features of CorC to check \IFbCOO information flow policies:

\begin{itemize}
	\item Programs are written in a tree structure of refining \IFbCOO tuples (see Figure~\ref{fig:refinement}). Besides the functional specification, variables are labeled with a type \T\ in the tuples. 
	\item Each \IFbCOO refinement rule is implemented in CorC. Consequently, functional correctness and security can be constructed simultaneously.
	\item The information flow checks according to the refinement rules are executed automatically after each refinement.
	\item Each CorC-program is uniquely mapped to a method in a SIFO class. A SIFO class contains methods and fields that are annotated with security labels and type modifiers.
	\item A properties view shows the type \T\ of each used variable in an \IFbCOO tuple. Violations of the information flow policy are explained in the view.
\end{itemize}



\subsection{Case Studies and Discussion}

The implementation of \IFbCOO in the tool CorC enables us to evaluate the feasibility of the security mechanism by successfully implementing three case studies~\cite{hall2005fundamental,thum2012family} from the literature and a novel one in CorC.
The case studies are also implemented and type-checked in SIFO to confirm that the case studies are secure.
The newly developed \textit{Database} case study represents a secure system that strictly separates databases of different security levels. \textit{Email}~\cite{hall2005fundamental}  ensures that encrypted emails cannot be decrypted by recipients without the matching key. \textit{Paycard} (\texttt{http://spl2go.cs.ovgu.de/projects/57}) and \textit{Banking}~\cite{thum2012family} simulate secure money transfer without leaking customer data.
The \textit{Database} case study uses four security levels, while the others (\textit{Email}, \textit{Banking}, and \textit{Paycard}) use two.

\begin{table}[tb]
	\restoretx
	\centering
	\scriptsize
	\begin{tabularx}{\linewidth}{XYYYY}
		\toprule
		\textbf{Name} & \textbf{\#Security Levels} & \textbf{\#Classes} & \textbf{\#Lines of Code} & \textbf{\#Methods in CorC} \\ \midrule\midrule
		Database      & 4                          & 6                  & 156                      & 2                          \\ \midrule
		Email~\cite{hall2005fundamental}         & 2                          & 9                  & 807                      & 15                         \\ \midrule
		Banking~\cite{thum2012family}       & 2                          & 3                  & 243                      & 6                          \\ \midrule
		Paycard       & 2                          & 3                  & 244                      & 5                          \\ \bottomrule
	\end{tabularx}
	\caption{Metrics of the case studies}
	\label{tab:evaluation}
\end{table}

As shown in Table~\ref{tab:evaluation}, the cases studies comprise three to nine classes with 156 to 807 lines of code each. 28 Methods that exceed the complexity of getter and setter are implemented in CorC. It should be noted that we do not have to implement every method in CorC. If only \low\ input data is used to compute \low\ output, the method is intrinsically secure. For example, three classes in the Database case study are implemented with only \low\ security levels. Only the class \texttt{GUI} and the main method of the case study, which calls the \low\ methods with higher security levels (using multiple method types) is then correctly implemented in CorC. The correct and secure promotion of security levels of methods called in the main method is confirmed by CorC.

\paragraph{Discussion and Applicability of IFbCOO.}

We emphasize that CbC and also \IFbCOO should be used to implement correctness- and security-critical programs~\cite{kourie2012correctness}. The scope of this work is to demonstrate the feasibility of the incremental construction of correctness- and security-critical programs. We argue that we achieve this goal by implementing four case studies in CorC. 

The constructive nature of \IFbCOO is an advantage in the secure creation of programs. Instead of writing complete methods to allow a static analyzer to accept/reject the method, with \IFbCOO, we directly design and construct secure methods. We get feedback during each refinement step, and
we can observe the status of all accessible variables at any time of the method. For example, we received direct feedback when we manipulated a \low\ object in the body of a \high\ then-branch. With this information, we could adjust the code to ensure security.
As \IFbCOO extends CorC, functional correctness is also guaranteed at the same time. This is beneficial as a program, which is security-critical, should also be functionally correct.
As \IFbCOO is based on SIFO, programs written with any of the two approaches can be used interchangeably. This allows developers to use their preferred environment to develop new systems, re-engineer their systems, or integrate secure software into existing systems.
These benefits of \IFbCOO are of course connected with functional and security specification effort, and the strict refinement-based construction of programs.


\section{Related Work}

In this section, we compare \IFbCOO to IFbC~\cite{runge2020lattice,schaefer2018towards} and other Hoare-style logics for information flow control. We also discuss information flow type systems and correctness-by-construction~\cite{kourie2012correctness} for functional correctness.

\IFbCOO extends IFbC~\cite{runge2020lattice} by introducing object-orientation and type modifiers. IFbC is based on a simple while language. As explained in Section~\ref{sec:formalizing}, the language of \IFbCOO includes objects and type modifiers. Therefore, the refinement rules of IFbC are revised to handle secure information flow with objects.
The object-orientation complicates the reasoning of secure assignments because objects could be altered through references with different security levels. If private information is introduced, an already public reference could read this information. SIFO and therefore \IFbCOO consider these cases and prevent information leaks by considering immutability and encapsulation and only allowing secure aliases.

Previous work using Hoare-style program logics with information flow control analyzes programs after construction, rather than guaranteeing security during construction. Andrews and Reitman~\cite{AndrewsR80} encode information flow directly in a logical form. They also support parallel programs. Amtoft and Banerjee~\cite{AmtoftB04} use Hoare-style program logics and abstract interpretations to detect information flow leaks. They can give error explanations based on strongest postcondition calculation. The work of Amtoft and Banerjee~\cite{AmtoftB04} is used in SPARK Ada~\cite{AmtoftHRRHG08} to specify and check the information flow.

Type system for information flow control are widely used, we refer to Sabelfeld and Myers~\cite{SabelfeldM03} for a comprehensive overview. We only discuss closely related type systems for object-oriented languages~\cite{barthe1999partial,banerjee2002secure,sun2004modular,Myers:1999:JPM:292540.292561,strecker2003formal,barthe2007certified}. Banerjee et al.~\cite{banerjee2002secure} introduced a type system for a Java-like language with only two security levels. We extend this by operating on any lattice of security levels. We also introduce type modifiers to simplify reasoning in cases where objects cannot be mutated or are encapsulated. Jif~\cite{Myers:1999:JPM:292540.292561} is a type system to check information flow in Java. One main difference is in the treatment of aliases: Jif does not have an alias analysis to reason about limited side effects. Therefore, Jif pessimistically discards programs that introduce aliases because Jif has no option to state immutable or encapsulated objects. \IFbCOO allows the introduction of secure aliases.

In the area of correctness-by-construction,
Morgan~\cite{morgan-book} and Back~\cite{back2012refinement} propose refinement-based approaches which refine functional specifications to concrete implementations. Beside of pre-/postcondition specification, Back also uses invariants as starting point. Morgan's calculus is implemented in ArcAngel~\cite{oliveira2003arcangel} with the verifier ProofPower~\cite{zeyda2009proofpower}, and SOCOS~\cite{back2009invariant,back2007testing} implements Back's approach. In comparison to \IFbCOO, those approaches do not reason about information flow security. Other refinement-based approaches are Event-B~\cite{EventB,abrial2010rodin} for automata-based systems and Circus~\cite{oliveira2009utp,oliveira2008crefine} for state-rich reactive systems. These approaches have a higher abstraction level, as they operate on abstract machines instead of source code. Hall and Chapman~\cite{hall2002correctness} introduced with CbyC another related approach that uses formal modeling techniques to analyze the development during all stages (architectural design, detailed design, code)  to eliminate defects early. \IFbCOO is tailored to source code and does not consider other development phases.

\section{Conclusion}
In this paper, we present \IFbCOO, which establishes an incremental refinement-based approach for functionally correct and secure programs.
With \IFbCOO programs are constructed stepwise to comply at all time with the security policy. The local check of each refinement can reduce debugging effort, since the user is not warned only after the implementation of a whole method. We formalized \IFbCOO by introducing 13 refinement rules and proved soundness by showing that constructed programs are well-typed in SIFO. We also implemented \IFbCOO in CorC and evaluated our implementation with a feasibility study.
One future direction of research is the conduction of comprehensive user studies for user-friendly improvements which is only now possible due to our sophisticated CorC tool support.


\paragraph*{Acknowledgments}
This work was supported by KASTEL Security Research Labs.

\balance
\bibliographystyle{splncs04}
\bibliography{Bibliography}

\appendix

\section{Expression Typing Rules of SIFO}
\label{sec:appendix_typing}

In Figure~\ref{f:typing}, typing rules of SIFO are shown.

\begin{figure}[!h]
	\centering
	\begin{scriptsize}
		\begin{mathpar}
			\inferrule
			{C \leq C'\\
				\mdf \leq \mdf'
			}{
				s\ \mdf\ C \leq s\ \mdf'\ C'}
			\quad (\textsc{Sub})
			
			\inferrule
			{\Gamma \vdash e: T'\\
				T' \leq T
			}{
				\Gamma \vdash e:T}
			\quad (\textsc{Subsumption})
			
			\inferrule
			{{}_{}}
			{\Gamma \vdash x : \Gamma(x)}
			\quad (\textsc{T-Var})
			
			\inferrule{
				\Gamma \vdash e_0 : s_0\ \mdf_0\ C_0\\
				s_1\ \mdf_1\ C_1\ f \in \fields(C_0)
			}{
				\Gamma \vdash e_0.f : \lub(s_0,s_1)\ \mdf_0\mdfarrow\mdf_1\ C_1
			}\quad (\textsc{Field Access})
			
			\inferrule
			{\Gamma \vdash e_0 : s_0\ \mut\ C_0 \\\\
				\Gamma \vdash e_1 : \lub(s_0,s)\ \mdf\ C\\
				s\ \mdf\ C\ f \in \fields(C_0)
			}
			{\Gamma \vdash e_0.f = e_1 : \lub(s_0,s)\ \mdf\ C}
			\quad (\textsc{Field Assign})
			
			\inferrule
			{\Gamma \vdash e_0 : T_0 \dots \Gamma \vdash e_n : T_n\\
				\security(T) \geq \security(T_0)\\\\
				\mathit{if\ }\mdf(T_i) \in \{\mut, \capsule\} 
				\mathit{\ then\ }
				\sec(T_i) \geq \sec(T_0)\\
				T_0 \dots T_n \rightarrow T \in \methTypes(\class(T_0),\ m)
			}
			{\Gamma \vdash e_0.m(e_1\dots e_n) : T}
			\quad (\textsc{Call})
			
			\inferrule
			{\Gamma \vdash e_1 : T_1[s] \dots \Gamma \vdash e_n : T_n[\s]\\
				\fields(C)=T_1 \ f_1\dots T_n\ f_n
			}
			{\Gamma \vdash \newKw\ s\ C(e_1\dots e_n) : \s\ \mut\ \C}
			\quad (\textsc{New})
			
			\inferrule
			{\Gamma[\mut\backslash\read] \vdash e : \s\ \mut\ \C
			}
			{\Gamma \vdash e: \s\ \capsule\ \C}
			\quad (\textsc{Prom})
			
			\inferrule{
				s' \leq s \\ \Gamma \vdash e : \s'\ \mdf\ C\\
				\mdf \in \{\imm,\capsule\}
			}{
				\Gamma \vdash e: s\ \mdf\ C}
			\quad (\textsc{Sec-Prom})
			
			\inferrule
			{
				\thisKw : s\ \mdf\ C,\ x_1 : T_1 \ldots x_n : T_n\\
				\vdash e : T
			}
			{C \vdash s\ \mdf\ \methodKw\ T\ m(T_1\ x_1\ldots T_n\ x_n) \{\returnKw\ e;\}}
			\quad (\textsc{M-Ok})
			
			\inferrule{
				C \vdash \MOne \dots C \vdash M_n\\
				\mathit{mhs}(\overline{C}) \subseteq \mathit{mhs}(\MOne\ldots M_n)
			}{
				\classKw\ C\ \implementsKw\ \overline{C}\ \{\overline{F}\ \MOne\ldots M_n\}
			}
			\quad (\textsc{C-Ok})
			
			\inferrule{
				\mathit{mhs}(\overline{C}) \subseteq \overline{\mathit{MH}}\\
			}{
				\interfaceKw\ C\ \extendsKw\ \overline{C}\ \{\overline{\mathit{MH}}\}
			}
			\quad (\textsc{I-Ok})
			
		\end{mathpar}
	\end{scriptsize}
	\caption{Expression typing rules of SIFO}
	\label{f:typing}
\end{figure}

\section{Method Declarations.}
\label{sec:methods}

In Figure~\ref{f:method_overloading}, we show the transformations, which can be applied on a method declaration.
In the first case, the security level of all types can be increased by choosing a security level $s'$ higher than the lowest level of the lattice. As used in the constructor rule, the helper function $T[s']$ increases the security level of $T$ to the least upper bound of $s'$ and $sec(T)$. This means, a method can be declared with some security level, but it can be used with any higher level by applying the first transformation case. For example, a mathematical function can be called with \low\ parameters and it returns a \low\ result, or it can be called with \high\ parameters and it returns a \high\ result. Thanks to this definition, the programmer has to declare the method only once.

In the second and third case, the security level can be altered as in the first case. Additionally, in the second case, all type modifiers \mut\ are changed to \capsule. This is useful for methods, which are declared with \mut\ parameters and return type.
If we provide \capsule\ instead of \mut\ in the input, we can use the method to produce a \capsule\ return value. If we ensure that only encapsulated input is used, we can guarantee that the produced output is also a \capsule. Thus, we do not have to declare a method twice with \capsule\ or \mut\ types. 

The third case changes all \read\ type modifiers to \imm\ and all \mut\ type modifiers to \capsule. This case is useful if the method is returning a \read\ value, then we can obtain an \imm. A \read\ reference can point to immutable or mutable object. If it is an immutable object, we can return it as \imm. If it is a mutable object, as in case two, the object can be promoted to \capsule, and \capsule\ is a subtype of \imm.

\begin{figure}[t]
	\begin{minipage}{0.97\linewidth}
		If $s\ \mdf\ \text{\lst@method@}\ T\ m (T_1\ x_1 \dots T_n\ x_n)$ is declared in $C$, with $T_0 = s\ \mdf\ C$ then\\
		1: $T_0[s'] \dots T_n[s'] \rightarrow T[s'] \in \methTypes(C, m)$\\
		2: $(T_0[s'] \dots T_n[s'] \rightarrow T[s'])[\mut\backslash \capsule] \in \methTypes(C, m)$\\
		3: $(T_0[s'] \dots T_n[s'] \rightarrow T[s'])[\read\backslash \imm, \mut\backslash\capsule] \in \methTypes(C, m)$\\
	\end{minipage}
	\caption{Definition of multiple method types}
	\label{f:method_overloading}
\end{figure}

\section{Proof of Soundness Theorem 1}
\label{sec:appendix_proof}

\begin{proof}
	We prove by structural induction on the length of the refinement chain. We prove that any expression that is generated using $n$ or less refinement steps is well-typed. 
	
	\noindent Base: The only expressions that can be generated with one
	refinement step are of form $x$ and $\mathtt{new} C()$. An expression of form $x$ must have been obtained starting from an abstract expression
	$\eA:[\Gamma;T]$ and was replaced with an $x$ so that $\Gamma(x)=T$. Thus, we can apply
	rule (T-Var) of SIFO to type the expression.
	
	 An expression of form $\mathtt{new} C()$ must have been obtained starting from an abstract expression
	 $\eA:[\Gamma;T]$ and was replaced with a $\mathtt{new} s C()$
	 so that $T= s\ \mut\ C$
	 and $\fields(C)=\emptyset$. Thus, we can apply rule (New) of SIFO to type the expression.
	
	\noindent Inductive step:
	For the inductive step, we prove by cases over the first performed refinement step.
	The expression resulting from the complete refinement will have
	sub-expressions that have been completed using less refinement steps.
	Thus, in all those cases the following inductive hypothesis will hold:
	all the sub-expressions of the final result of the refinement using less than $n$ refinement steps are well-typed. We will then need to prove that the final result of the refinement is well-typed too.
	
	\textbf{Field Assignment}. The abstract expression
	$\eA:[\Gamma;T]$ is refined in the first step into $\eA_0.f=\eA_1$
	and then it is fully-refined into $e_0.f=e_1$.
	It must be the case that
	$\eA_0:[..]$ is fully-refined into $e_0$
	(by application of refinement rules through the rule 9 (context rule))
	and $\eA_1:[..]$ is fully-refined into $e_1$
	(by application of refinement rules through the rule 9 (context rule)).
	By the inductive hypothesis, we have
	$\Gamma \vdash e_0:s_0\ \mut\ C_0$ and  $\Gamma \vdash e_1:s_1\ \mdf\ C$.
	Then, we know that $\Gamma \vdash e_0.f=e_1 : s_1\ \mdf\ C$,
	since we applied refinement rule (Field Assignment). We
	know that we can apply the rule (Field Assign) of SIFO because
	premise 1 of (Field Assign) is satisfied since $e_0$ is well-typed and
	premise 2 of (Field Assign) is satisfied since $e_1$ is well-typed. Additionally, the
    (Field Assignment) refinement states that $\s\ \mdf\ \C\ f \in fields(C_0)$  and $s_1 = lub(s_0,s)$. These are the same conditions as for the (Field Assign) typing rule in SIFO, therefore, the (Field Assignment) refinement is well-typed.
	
	\textbf{Field Access}.
	The abstract expression
	$\eA:[\Gamma;T]$ is refined in the first step into $\eA_0.f$
	and then it is fully-refined into $e_0.f$.
	It must be the case that
	$\eA_0:[..]$ is fully-refined into $e_0$.
	By the inductive hypothesis, we have
	$\Gamma \vdash e_0:s_0\ \mdf_0\ \C_0$.
	Then, we know that $\Gamma \vdash e_0.f : s\ \mdf\ C$ and $\s_1\ \mdf_1\ \C\ f \in fields(C_0)$,
	since we applied refinement rule (Field Access). We
	know that we can apply the rule (Field Access) of SIFO because
	premise 1 of (Field Access) is satisfied since $e_0$ is well-typed. Additionally, the
	(Field Access) refinement states that
	and $s = \lub(s_0,s_1)$ and $\mdf_0 \mdfarrow \mdf_1 = \mdf$.
	These are the same conditions as for the (Field Access) typing rule in SIFO, therefore, the (Field Access) refinement is well-typed.	
	
	\textbf{Method Call}. The abstract expression
	$\eA:[\Gamma;T]$ is refined in the first step into $\eA_0.m(\eA_1,\dots, \eA_n)$
	and then it is fully-refined into $e_0.m(e_1,\dots, e_n)$.
	It must be the case that
	$\eA_0:[..], \dots, \eA_n:[..]$ are fully-refined into $e_0, \dots, e_n$.
	By the inductive hypothesis, we have $\Gamma \vdash e_0 : T_0 \dots \Gamma \vdash e_n : T_n$.
	Then, we know that $\Gamma \vdash e_0.m(e_1,\dots, e_n) : T$,
	since we applied refinement rule (Method Call). We
	know that we can apply the rule (Call) of SIFO because
	premise 1 of (Call) is satisfied since $e_0, \dots, e_n$ are well-typed. Additionally, the (Method Call) refinement state that $T_0 \dots T_n \rightarrow T \in \methTypes(\class(T_0),\ m)$ and 
	$\security(T) \geq \security(T_0)$ and if $mdf(T_i) \in \{\mut,\capsule\}$ then $\security(T_i) \geq \security(T_0)$. These are the same conditions as for the (Call) typing rule, therefore, the (Method Call) refinement is well-typed.

	\textbf{Constructor}.  The abstract expression
	$\eA:[\Gamma;T]$ is refined in the first step into $\mathtt{new}\ s\ C(\eA_1\dots \eA_n)$
	and then it is fully-refined into $\mathtt{new}\ s\ C(e_1\dots e_n)$.
	It must be the case that
	$\eA_1:[..], \dots, \eA_n:[..]$ are fully-refined into $e_1, \dots, e_n$.
	By the inductive hypothesis, we have $\Gamma \vdash e_1 : T_1[s] \dots \Gamma \vdash e_n : T_n[s]$.
	Then, we know that $\mathtt{new}\ s\ C(\eA_1\dots \eA_n)$,
	since we applied refinement rule (Constructor). We
	know that we can apply the rule (New) of SIFO because
	premise 1 of (New) is satisfied since $e_0, \dots, e_n$ are well-typed. Additionally, the (Constructor) refinement state that $\fields(C)=T_1 \ f_1\dots T_n\ f_n$. This is the same condition as for the (New) typing rule in SIFO, therefore, the (Constructor) refinement is well-typed.
	
	

	
	\textbf{Subsumption}. The abstract expression 
	$\eA:[\Gamma;T]$  is refined in the first step into $\eA_1:[\Gamma;T']$
	and then it is fully-refined into $e$. By the inductive hypothesis, we have $\Gamma \vdash e : T'$.
	Then, we know that $T' \leq T$,
	since we applied refinement rule (Subsumption).
	We know that we can apply the rule (Subsumption) of SIFO because
	premise 1 and 2 are satisfied. Therefore, the (Subsumption) refinement is well-typed.

	\textbf{Security Promotion}. 
	The abstract expression $\eA:[\Gamma;T]$  is refined in the first step into $\eA_1:[\Gamma;s'\ \mdf\ C]$
	and then it is fully-refined into $e$. By the inductive hypothesis, we have $\Gamma \vdash e : s'\ \mdf\ C$.
	Then, we know that $s' \leq s$ and $\mdf \in \{\capsule,\imm\}$
	since we applied refinement rule (Security Promotion).
	We know that we can apply the rule (Sec-Prom) of SIFO because
	premise 1, 2, and 3 are satisfied. Therefore, the (Security Promotion) refinement is well-typed.

	\textbf{Modifier Promotion}.The abstract expression $\eA:[\Gamma;s\ \capsule\ C]$  is refined in the first step into $\eA_1:[\Gamma[\mut\backslash\read];s\ \mut\ C]$
	and then it is fully-refined into $e$. By the inductive hypothesis, we have $\Gamma[\mut\backslash\read] \vdash e : s\ \mut\ C$.
	We know that we can apply the rule (Prom) of SIFO because
	premise 1 is satisfied. Therefore, the (Modifier Promotion) refinement is well-typed.

\end{proof}

\end{document}